\documentclass{sigchi-ext}
\usepackage[T1]{fontenc}
\usepackage{textcomp}
\usepackage[scaled=.92]{helvet} 
\usepackage{graphicx} 
\usepackage{balance}  
\usepackage{booktabs} 
\usepackage{ccicons}  
\usepackage{ragged2e} 
\usepackage{cite}

\def\plaintitle{Human Evaluation of Interpretability: The Case of AI-Generated Music Knowledge}

\def\emptyauthor{}
\def\plainkeywords{Interpretable AI and machine learning; interpretation of AI-generated knowledge/rules; human evaluation; assessment on verbal materials}

\title{\plaintitle}

\numberofauthors{6}
\author{%
  \alignauthor{%
    \textbf{Haizi Yu}\\
    \vspace{0.04in}
    \affaddr{Coordinated Science Lab} \\
    \affaddr{University of Illinois at Urbana-Champaign} \\
    \vspace{0.04in}
    \affaddr{Knowledge Lab} \\
    \affaddr{University of Chicago} \\
    \vspace{0.04in}
    \email{haiziyu7@illinois.edu} }\vspace{0.07in}
  \alignauthor{%
    \textbf{Lav R. Varshney}\\
    \vspace{0.04in}
    \affaddr{Coordinated Science Lab} \\
    \affaddr{University of Illinois at Urbana-Champaign} \\
    \vspace{0.04in}
    \affaddr{Salesforce Research} \\
    \affaddr{Palo Alto, CA} \\
    \vspace{0.04in}
    \email{varshney@illinois.edu} }
  \vfil
  \alignauthor{%
    \textbf{Heinrich Taube}\\
    \vspace{0.04in}
    \affaddr{School of Music}\\
    \affaddr{University of Illinois at Urbana-Champaign}\\
    \vspace{0.04in}
    \email{taube@illinois.edu} }\vspace{0.07in}
  \vfil
  \alignauthor{%
    \textbf{James A. Evans}\\
    \vspace{0.04in}
    \affaddr{Knowledge Lab}\\
    \affaddr{University of Chicago} \\
    \vspace{0.04in}
    \email{jevans@uchicago.edu} }
}

\definecolor{linkColor}{RGB}{6,125,233}
\hypersetup{%
  pdftitle={\plaintitle},
  pdfauthor={\emptyauthor},
  pdfkeywords={\plainkeywords},
  bookmarksnumbered,
  pdfstartview={FitH},
  colorlinks,
  citecolor=black,
  filecolor=black,
  linkcolor=black,
  urlcolor=linkColor,
  breaklinks=true,
}

\newcommand{\ie}{i.e.\ }
\newcommand{\eg}{e.g.\ }

\newcommand{\RN}[1]{\textup{\uppercase\expandafter{\romannumeral#1}}}

\begin{document}

\CopyrightYear{2020}
\setcopyright{rightsretained}
\conferenceinfo{CHI'20,}{April  25--30, 2020, Honolulu, HI, USA}
\isbn{978-1-4503-6819-3/20/04}
\doi{https://doi.org/10.1145/3334480.XXXXXXX}
\copyrightinfo{\acmcopyright}

\maketitle

\RaggedRight{} 

\begin{abstract}
Interpretability of machine learning models has gained more and more attention among researchers in the artificial intelligence (AI) and human-computer interaction (HCI) communities.
Most existing work focuses on decision making, whereas we consider knowledge discovery.
In particular, we focus on evaluating AI-discovered knowledge/rules in the arts and humanities.
From a specific scenario, we present an experimental procedure to collect and assess human-generated verbal interpretations of AI-generated music theory/rules rendered as sophisticated symbolic/numeric objects.
Our goal is to reveal both the possibilities and the challenges in such a process of decoding expressive messages from AI sources.
We treat this as a first step towards 1) better design of AI representations that are human interpretable and 2) a general methodology to evaluate interpretability of AI-discovered knowledge representations.
\end{abstract}



Please use the 2012 Classifiers and see this link to embed them in the text: \url{https://dl.acm.org/ccs/ccs_flat.cfm}

\section{Introduction}
Machine-learning (ML) and data-driven models in general have achieved tremendous success in recent decades.
Yet, their interpretability does not scale with their superior powers in learning complex patterns and fast-growing accuracy in prediction ~\cite{Molnar2019,DoshiK2017}.
In fact, some even argue there is a trade off between model performance and interpretability~\cite{KuhnJ2013}.
As a result, many ML models---especially powerful deep neural networks---are often branded as \emph{black boxes}, and there has been a growing demand from both the ML and the HCI community pressing for interpretable ML that are sensible to people. 

\begin{marginfigure}[0pc]
  \begin{minipage}{\marginparwidth}
    \centering
    \includegraphics[width=\marginparwidth]{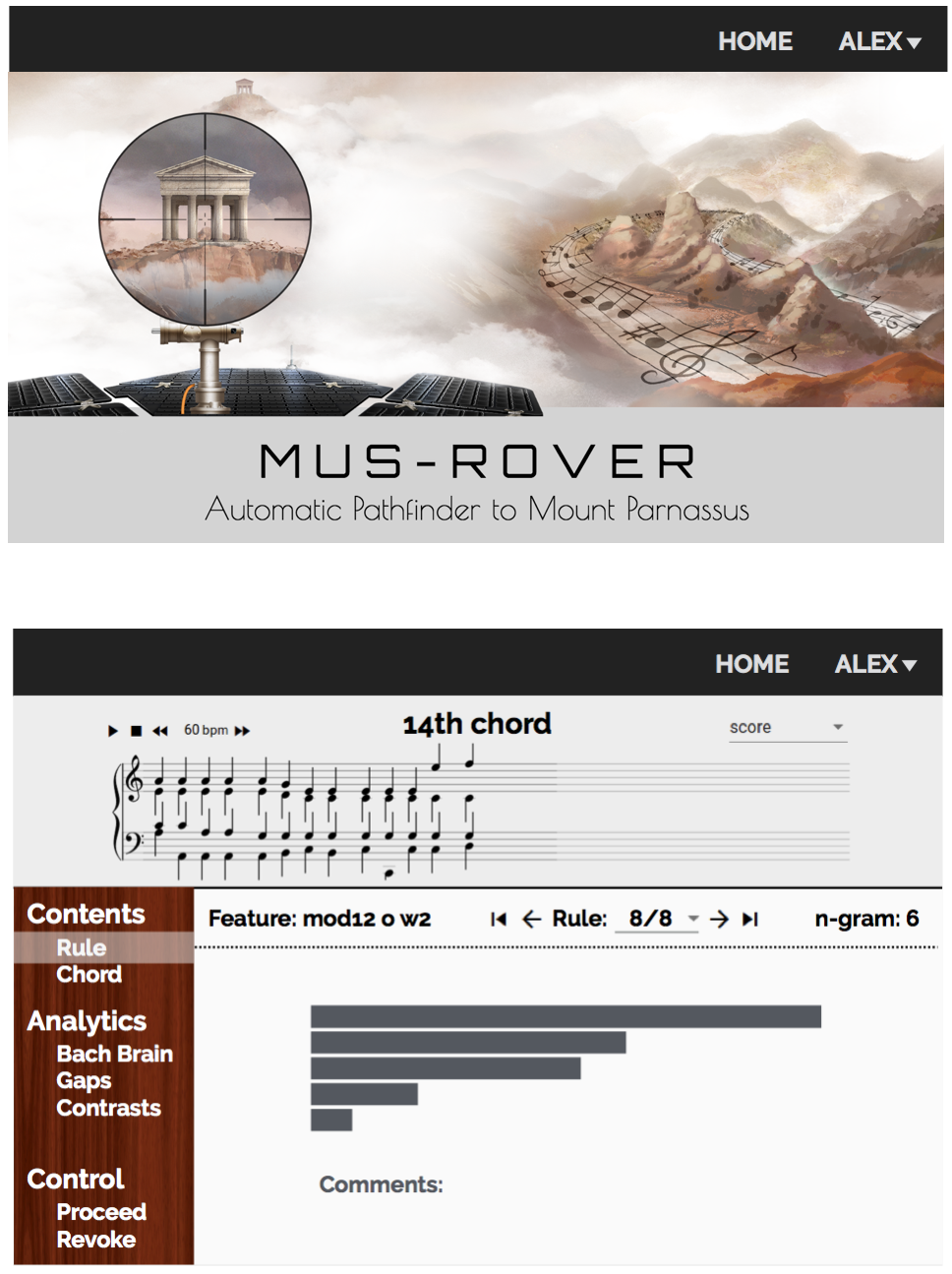}
    \caption{The MUS-ROVER web application: an automatic music theorist and pedagogue that offers people rules on music composition.}~\label{fig:mus-rover-webapp}
  \end{minipage}
\end{marginfigure}

The term interpretability has not yet been precisely defined in any rigorous sense~\cite{MurdochSKAY2019,DoshiK2017}, even though scholars have recruited widely shared intuition to suggest several broad, possible meanings.
Some works attempt to define ML interpretability or ML explainability in general, while others list criteria for an AI system to be considered interpretable, and still others propose forms, methods, or metrics to demonstrate interpretability from a certain perspective (\eg visualization, mathematical equations, representative examples).
Nevertheless, there remains a long way to go to reach formal sufficient and necessary conditions for interpretability.
Researchers also argue that interpretability should be defined separately in different contexts, \eg the interpretability of \emph{what} (\eg model output or the model itself) and interpretable to \emph{whom} (\eg model developers, domain experts, or the general crowd).
Further, most existing work on ML interpretability has focused on decisions, rather than complicated, discovered knowledge as here.

In this paper, we present an experimental procedure to evaluate interpretability of the output of an AI system that discovers knowledge.
The system, called MUS-ROVER, is a data-driven model we developed to learn music composition rules and concepts (output) from sheet music (input).
We designed an experiment to collect and assess human written interpretations of AI-generated music rules in order to assess how interpretable the rules themselves are to humans.
One main challenge here is that our collected human interpretations are verbal (like comments) instead of numerical (like ratings) or symbolic (like answers from multiple-choice questions).
Our experiment is specifically designed for our Music AI project, and it focuses on only interpretability of the model output.
It does not test the interpretability of the model itself, nor does it test performance of the model output (\eg rule expressiveness).
However, from this case study, we aim to take a first step towards a systematic and commonly-accepted procedure that can be broadly applied to ML models in various topic domains, especially when human feedback is in free form text.

\section{MUS-ROVER: the AI System Overview}

MUS-ROVER is a self-rule-learning system that learns music composition rules from sheet music  (Figure~\ref{fig:mus-rover-webapp}).
It plays the role of both an automatic music theorist and an automatic music pedagogue, but not an automatic composer.
It does not output music, but instead rules to teach people music composition in a given style.

Every rule is formulated as a histogram, tracing an empirical probability distribution that abstractly represents musical chords.
An $n$-note chord is mathematically represented by an $n$-dimensional vector of MIDI numbers, with each dimension denoting a voice (\eg soprano, bass). An abstraction of chords is mathematically represented by an equivalence relation on the chord space (\eg identifying chords (C,E,G) and (F,A,C) as equivalent by the fact that they are both major triads).
Here, we focus on feature-induced abstractions, meaning every equivalence relation is induced from a feature function that maps equivalent chords to the same feature value.
So, given a feature function, a rule is a histogram of all possible feature values.
In short, there are two pieces of information: feature and probability.

\begin{marginfigure}[2pc]
  \begin{minipage}{\marginparwidth}
    \centering
    \includegraphics[width=\marginparwidth]{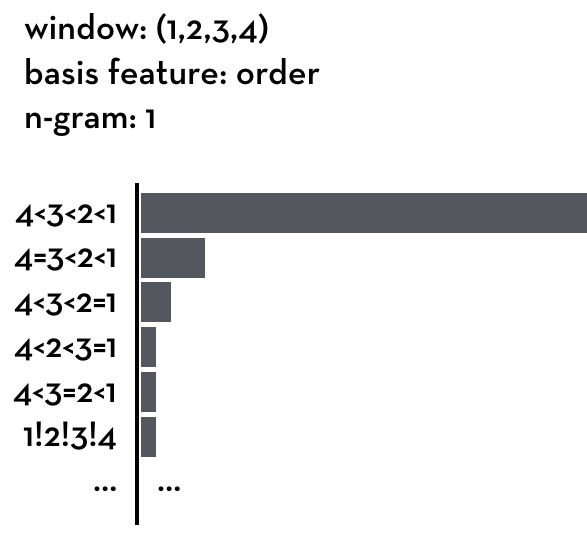}
    \caption{Example of a rule from MUS-ROVER. The four dimensions correspond to four human vocal voices: $1$-soprano, $2$-alto, $3$-tenor, $4$-bass. The window function extracts all four dimensions (\ie an identity function); the basis feature function is the so-called order function that represents a better-designed argsort function to handle ties ($=$) and incomparability ($!$) in the presence of a music rest. Possible feature values are listed as $y$-labels in the histogram. There is a dominating probability peak located at feature value $4<3<2<1$. This rule describes the relative highness among the four voices, and the histogram shows that  a higher voice almost always sings a higher music pitch.}~\label{fig:mus-rover-rule}
  \end{minipage}
\end{marginfigure}

Both the feature and the distribution of a rule have another layer of complexity.
First, every feature function is systematically generated as the composition of a window function followed by one or more basis feature functions.
The window function selects a subset of dimensions out of a chord vector (\ie selecting musical voices), and each basis feature function is a simple arithmetic operator (\eg subtraction, sort, modulo).
Second, every probability distribution comes from a family of $n$-gram probability distributions.
For example, a $1$-gram probability distribution is an unconditional probability distribution, and a $k$-gram (for $k>1$) probability distribution is a conditional probability distribution conditioned on the feature values of the previous $(k-1)$ chords.
To summarize, for one to interpret a rule, (s)he is expected to assemble the following information:
\vspace{-0.15in}
\begin{itemize}\setlength\itemsep{0in}
\item feature: window and basis feature(s);
\item probability: probability measure and conditional(s).
\end{itemize}
\vspace{-0.15in}
Figure~\ref{fig:mus-rover-rule} provides an example of a rule and one way of interpreting it reminiscent of avoiding voice crossing.

\section{Experiment}

Here we detail our designed experiment, the goal of which is to test to what degree our AI-generated rules are interpretable.
We target people who have entry-level math and music theory knowledge; \ie by interpretability we mean interpretable to them.
The whole experimental procedure is divided into two phases: 1) collecting human interpretations of the rules generated by MUS-ROVER and 2) assessing collected human interpretations to further evaluate the interpretability of AI-generated rules.

\subsection{Collect Human Interpretations}

The experiment was conducted in the form of a two-week written homework assignment for 23 students.
Students came from the CS+Music degree program recently launched at the University of Illinois at Urbana-Champaign.
Entry-level knowledge of computer science, related math, and music theory is assumed from every student.
However, all students are new to the MUS-ROVER system, and none has read any rules from MUS-ROVER before.

The homework contained three parts.
Part~\RN{1} includes detailed instructions on the format of the rules including both feature-related and probability-related instructions respectively.
More specifically, we provide verbal definition, mathematical representation, and typical examples for each of the following terms: chord, window, basis feature, feature, rule, $n$-gram, histogram, data set.
A faithful understanding of these eight terms is the only prerequisite to complete the homework.
The estimated reading time of instructions is about an hour.
Once this self-pretraining phase is completed, the students are ready to go to the second and third parts---the main body of the homework.

Part~\RN{2} contains 11 $1$-gram rules---a histogram specified by window and basis feature(s);
Part~\RN{3} contains 14 $2$-gram rules---a histogram now specified by window, basis feature(s), and a conditional.
We asked the students to freely write what they see in each of the histograms guided by the following two questions: 1) does the histogram agree/disagree with any of the music concepts/rules you know (write done the music concepts/rules in music-theoretic terms)? 2) does the histogram suggest something new (\ie neither an agreement nor a disagreement, with no clear connection to any known knowledge)?
Answers to each of the $25$ rules comes in the form of text (essay style), containing word descriptions that ``decode'' the histogram---a symbolic and pictorial encoding.
Students were explicitly instructed that writing out a description that involves a literal repetition of the histogram (\eg taking a modulo $12$ of a chord results in a $90\%$ chance of being $(0,0,4,7)$) is not acceptable: they must reveal the music behind the math.
In fact, we only want qualitative descriptions. 
Students were specifically told (in the instructions) to only pay attention to relative values of the probabilities whose exact numbers are unimportant (\eg what are most likely, what is more likely, and/or what are almost impossible).

\begin{margintable}[-5pc]
  \begin{minipage}{\marginparwidth}
    \centering
    \begin{tabular}{r|r}
      {\small \textbf{Score Range}} & {\small \textbf{\# of Students}} \\
      \toprule
      50 & 3   \\
      $[$40,50$)$ & 7  \\
      $[$30,40$)$ & 2  \\
      $[$20,30$)$ & 4 \\
      $[$10,20$)$ & 1 \\
      $[$0,10$)$ & 1 \\
      0 & 5
    \end{tabular}
    \caption{Students' final scores.}~\label{tab:result}
  \end{minipage}
\end{margintable}

This homework was due two weeks after initially released.
During the two-week period, we asked students to complete it independently (\eg no group studies or office hours).

\subsection{Assess Human Interpretations}

The homework was designed in a way such that every histogram rule encodes at least one music concept/rule consistent with standard music theory. In addition, every histogram contains either one additional known music rule or something strange that conflicts with a known rule or represents something new.
We assigned two points per rule.
Further, we made an initial rubric containing the (authoritative) music keywords used to describe every rule histogram.

Because students' answers arrived in the form of qualitative text, to ensure credibility and fairness of the initial rubric, we held a discussion session at a regular lecture time (80 minutes) with all students as well as the teaching staff.
During the discussion session, we went over all 25 rules one by one.
For each, we first announced the keywords in the initial rubric and explained to the students that these keywords would later be used to grade their homework.
However, in the discussion session, every student was encouraged to object to any of our announced keywords and/or to propose new keywords accompanied with a convincing explanation.
New/modified keywords that were commonly agreed upon were added/updated to the initial rubric. 
By the end of discussion session, we compiled a more inclusive rubric containing broadly accepted keywords.
This rubric-generating process was transparent to all the students.

In the final step, we manually graded every student's answer sheet against keywords in the rubric and computed their scores.
A summary of the students' performances is presented in Table~\ref{tab:result}.
Except for cases where the student did not do the homework, a major source of score deduction was from misunderstanding the $n$-gram (\eg the probability of the current chord conditioned on the previous chord was mistakenly interpreted as the probability of the previous chord conditioned on the current one).
This is largely due to unfamiliarity with the $n$-gram models for new CS+Music students.
Nevertheless, the majority of the students who did the homework succeeded in interpreting the rules generated from an AI system, which in turn provides evidence on the interpretability of AI output itself.

\balance{} 

\bibliographystyle{SIGCHI-Reference-Format}
\bibliography{sample}

\end{document}